\documentclass[onecolumn,unpublished,a4paper]{quantumarticle}
\usepackage[utf8]{inputenc}
\usepackage[backend=biber,style=alphabetic,maxbibnames=999]{biblatex}
\usepackage{amsmath}
\usepackage{amssymb}
\usepackage{amsthm}
\usepackage{amsfonts}
\usepackage[caption=false]{subfig}
\usepackage[colorlinks]{hyperref}
\usepackage[all]{hypcap}
\usepackage{tikz}
\usepackage{relsize}
\usepackage{color,soul}
\usepackage{capt-of}
\usepackage{mathtools}
\usepackage{float}
\usepackage[section]{placeins}
\usepackage{listings}
\usepackage[T1]{fontenc}  
\usetikzlibrary{decorations.pathreplacing}
\usepackage{braket}

\DeclareFixedFont{\ttb}{T1}{txtt}{bx}{n}{4}
\DeclareFixedFont{\ttm}{T1}{txtt}{m}{n}{4}
\definecolor{deepblue}{rgb}{0,0,0.5}
\definecolor{deepred}{rgb}{0.6,0,0}
\definecolor{deepgreen}{rgb}{0,0.5,0}
\newcommand\cppstyle{\lstset{
language=C++,
basicstyle=\ttm,
otherkeywords={uint8_t, __m256i, size_t, ASSERT_TRUE, EXPECT_TRUE, TEST, BENCHMARK},
keywordstyle=\ttb\color{deepblue},
emphstyle=\ttb\color{deepblue},
stringstyle=\color{deepgreen},
commentstyle=\fontfamily{txtt}\selectfont\color{gray},
showstringspaces=false,
literate={*}{{\char42}}1
         {-}{{\char45}}1
}}
\lstnewenvironment{cpp}[1][]
{\cppstyle\lstset{#1}}{}

\newcommand\pythonstyle{\lstset{
language=python,
basicstyle=\ttm,
morekeywords={assert,as,echo},
keywordstyle=\ttb\color{deepblue},
emphstyle=\ttb\color{deepblue},
stringstyle=\color{deepgreen},
commentstyle=\fontfamily{txtt}\selectfont\color{gray},
showstringspaces=false,
literate={*}{{\char42}}1
         {-}{{\char45}}1
}}
\lstnewenvironment{python}[1][]
{\pythonstyle\lstset{#1}}{}

\lstdefinestyle{stimcircuit}{
    language=python,
    basicstyle=\fontsize{6}{6}\selectfont\ttfamily,
    upquote=true,
    stepnumber=1,
    numbersep=8pt,
    showstringspaces=false,
    breaklines=true,
    frame=single,
    aboveskip=1.5em,
    belowskip=1.5em,
    commentstyle=\color{gray},
    classoffset=1,
    morekeywords={DETECTOR,OBSERVABLE_INCLUDE,rec},
    keywordstyle=\color{deepgreen},
    classoffset=2,
    morekeywords={H,R,MPP,M,RX,RY,MY,MX,SQRT\_X,XCY,XCZ,YCX},
    keywordstyle=\color{deepblue},
    classoffset=3,
    morekeywords={X_ERROR,DEPOLARIZE2,DEPOLARIZE1},
    keywordstyle=\color{red},
    classoffset=4,
    morekeywords={TICK,SHIFT_COORDS,QUBIT_COORDS},
    keywordstyle=\color{gray}
}

\hypersetup{linkcolor=blue}

\renewcommand\thesection{\arabic{section}}

\theoremstyle{definition}

\theoremstyle{definition}

\theoremstyle{definition}

\newcommand{\eq}[1]{\hyperref[eq:#1]{Equation~\ref*{eq:#1}}}
\renewcommand{\sec}[1]{\hyperref[sec:#1]{Section~\ref*{sec:#1}}}
\DeclareRobustCommand{\app}[1]{\hyperref[app:#1]{Appendix~\ref*{app:#1}}}
\newcommand{\fig}[1]{\hyperref[fig:#1]{Figure~\ref*{fig:#1}}}
\newcommand{\tbl}[1]{\hyperref[tbl:#1]{Table~\ref*{tbl:#1}}}
\newcommand{\theoremref}[1]{\hyperref[theorem:#1]{Theorem~\ref*{theorem:#1}}}
\newcommand{\definitionref}[1]{\hyperref[definition:#1]{Definition~\ref*{definition:#1}}}

\begin{document}
\title{Tetrationally Compact Entanglement Purification}

\date{\today}
\author{Craig Gidney}
\email{craig.gidney@gmail.com}
\affiliation{Google Quantum AI, Santa Barbara, California 93117, USA}

\begin{abstract}
This paper shows that entanglement can be purified using very little storage, assuming the only source of noise is in the quantum channel being used to share the entanglement.
Entangled pairs with a target infidelity of $\epsilon$ can be created in $\Tilde{O}(\log \frac{1}{\epsilon})$ time using $O(\log^{\ast} \frac{1}{\epsilon})$ storage space, where $\log^{\ast}$ is the iterated logarithm.
This is achieved by using multiple stages of error detection, with boosting within each stage.
For example, the paper shows that 11 qubits of noiseless storage is enough to turn entanglement with an infidelity of $1/3$ into entanglement with an infidelity of $10^{-1000000000000000000000000000}$.
\end{abstract}

\textbf{Data availability}: All code written for this paper is included in its attached ancillary files.

\tableofcontents

\section{Introduction}
\label{sec:introduction}

If Alice and Bob are linked by a noisy quantum channel, they can use error correction to reduce the noise of the channel.
For example, Alice can encode a qubit she wants to send into the $[[5,1,3]]$ perfect code~\cite{laflamma1996perfectcode}, before transmitting over the channel to Bob.
This increases the amount of data transmitted, but allows the message to be recovered if one error happens.
However, directly encoding the data to be sent is suboptimal.
A more effective technique is to use entanglement purification~\cite{bennett1995eprpurify}.

In an entanglement purification protocol, Alice never exposes important data to the noisy quantum channel.
The channel is only used to share EPR pairs (copies of the state $|00\rangle + |11\rangle$).
Once an EPR pair has been successfully shared, the message is moved from Alice to Bob using quantum teleportation~\cite{bennett1993teleportation}.
The benefit of doing this is that, if a transmission error happens when sharing the EPR pair, nothing important has been damaged.
EPR pairs are fungible; broken ones can be discarded and replaced with new ones.
This allows error correction to be done with error detection, which is more efficient and more flexible.
For example, discarding and retransmitting allows a distance three code to correct two errors instead of one.

The time complexity of entanglement purification is simple.
Starting from a channel with $O(1)$ infidelity, creating at least one EPR pair with a target infidelity of $\epsilon$ requires $\Theta(\log \frac{1}{\epsilon})$ uses of the noisy channel.
This time complexity is achievable with a variety of techniques, like encoding the entanglement into a quantum error correcting code with linear distance~\cite{panteleev2021goodldpc}.
This time complexity is optimal because, if fewer than $\Theta(\log \frac{1}{\epsilon})$ noisy pairs are shared, the chance of every single physical pair being corrupted would be larger than $\epsilon$.

Surprisingly, the space complexity of entanglement purification is not so simple.
My own initial intuition, which I think is shared by some other researchers based on a few conversions I've had, is that the space complexity would also be $\Theta(\log \frac{1}{\epsilon})$.
If storage was noisy then this would be the case, since otherwise the chance of every single storage qubit failing simultaneously would be greater than the desired error rate.
However, if you assume the only source of noise is the noisy channel (meaning storage is noiseless, local operations are noiseless, and classical communication is noiseless), then the space complexity drops dramatically even if you still require the time complexity to be $\Tilde{\Theta}(\log \frac{1}{\epsilon})$.

This paper explains a construction that improves entanglement fidelity \emph{tetrationally} versus storage space, achieving a space complexity of $O(\log^\ast \frac{1}{\epsilon})$.
It relies heavily on the assumption of perfect storage, which is unrealistic, but the consequences are surprising enough to merit a short paper.
Also, some of the underlying ideas are applicable to practical scenarios.

\section{Construction}
\label{sec:construction}

\subsection{Noise Model}

I'll be using a digitized noise model of shared EPR pairs.
Errors are modelled by assuming that, when an EPR pair is shared, an unwanted Pauli may have been applied to one of its qubits.
There are four Paulis that may be applied: the identity Pauli $I$ resulting in the correct shared state $B_I = |00\rangle + |11\rangle$, the bit flip $X$ resulting in the state $B_X = |01\rangle + |10\rangle$, the phase flip $Z$ resulting in the state $B_Z = |00\rangle - |11\rangle$, and the combined bit phase flip $Y$ resulting in the state $B_Y = |01\rangle - |10\rangle$.

I will represent noisy EPR pairs using a four dimensional vector of odds. The vector

$$\begin{bmatrix}w \\ x \\ y \\ z\end{bmatrix}$$

describes a noisy EPR pair where the odds of the applied Pauli being $I$, $X$, $Y$, or $Z$ are $w:x:y:z$ respectively.
Note that this is a degenerate representation: there are multiple ways to represent the same noisy state.
The exact density matrix of a noisy state described by an odds vector is:

\begin{equation}
\text{ErrModelDensityMatrix}\left(\begin{bmatrix}w\\x\\y\\z\end{bmatrix}\right) = \frac{w|B_I\rangle\langle B_I| + x|B_X\rangle\langle B_X| + y|B_Y\rangle\langle B_Y| + z|B_Z\rangle\langle B_Z|}{w+x+y+z}
\end{equation}

Note that this noise model can't represent coherent errors, like an unwanted $\sqrt[9]{X}$ gate being applied to one of the qubits of an EPR pair.
Some readers may worry that this means my analysis won't apply to coherent noise.
This isn't actually a problem, because the purification process will digitize the noise.
But, for the truly paranoid, coherent noise can be forcibly transformed into incoherent noise by Pauli twirling.
If Alice and Bob have an EPR pair that has undergone coherent noise, they can pick a Pauli $P \in \{I,X,Y,Z\}$ uniformly at random then apply $P$ to both qubits and forget $P$.
The density matrix describing the state after applying the random $P$, not conditioned on which $P$ was used, is expressible in the digitized noise model I'm using.
Discarding coherence information in this way is suboptimal, but sufficient for correctness.

Often, the exact odds will be inconveniently complicated and it will be beneficial to simplify them even if that makes the state worse.
For that purpose, I define $u \xrightarrow{\text{decay}} v$ to mean "a state with odds vector $u$ can be turned into a state with odds vector $v$ via the addition of artificial noise".
A sufficient condition for this relationship to hold is for the identity term to not grow and for the error terms to not shrink:

\begin{equation}
\Big(w_1 \geq w_2 \land x_1 \leq x_2 \land y_1 \leq y_2 \land z_1 \leq z_2\Big)
\implies
\left(\begin{bmatrix}w_1\\x_1\\y_1\\z_1\end{bmatrix}
\xrightarrow{\text{decay}}
\begin{bmatrix}w_2\\x_2\\y_2\\z_2\end{bmatrix}\right)
\end{equation}

\subsection{Distilling with Rep Codes}

\begin{figure}
    \centering
    \resizebox{\linewidth}{!}{
    \includegraphics{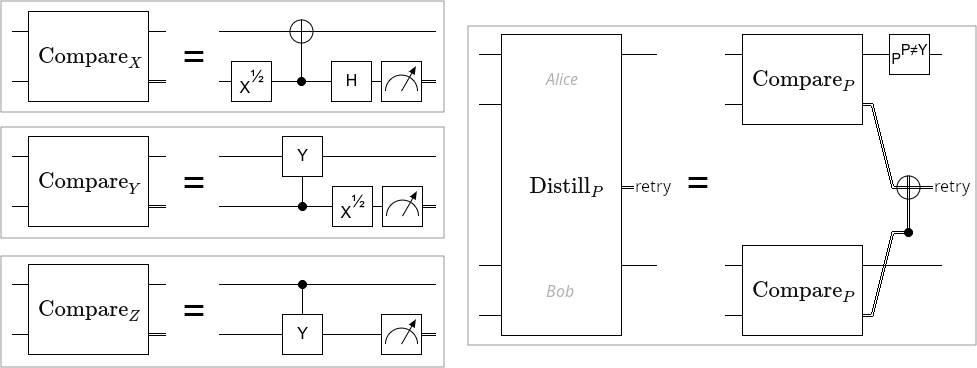}
    }
    \caption{
        The operations executed when distilling with a rep code in each basis.
        Each $\text{Compare}_P$ circuit has a measurement flow $+PP \rightarrow (-1)^m$ as well as decoding flows $X_L \rightarrow XI$ and $Z_L \rightarrow ZI$.
        The $P^{P\neq Y}$ gate removes minus signs coming from the fact that $|00\rangle + |11\rangle$ is stabilized by $-YY$ instead of $+YY$.
        It applies $X$ when $P=X$, $I$ when $P=Y$, and $Z$ when $P=Z$.
        Note that these circuits are somewhat non-standard, due to the underlying definitions of $X_L$ and $Z_L$ being somewhat non-standard, to control the exact way in which input sign errors propagate into output sign errors.
        The $\text{Distill}_P$ circuit has Alice and Bob each locally run $\text{Compare}_P$, and then use classical communication to compare the results.
        If the results agree, they keep the output qubits.
        Otherwise they discard the output qubits and try again with new inputs.
    }
    \label{fig:distill_blocks}
\end{figure}

The basic building block used by this paper is distillation with distance 2 rep codes.
There are three relevant rep codes: X basis, Y basis, and Z basis.
The X basis rep code has the stabilizer $XX$, the logical X observable $XI$, and the logical Z observable $ZY$.
The Y basis rep code has the stabilizer $YY$, the logical X observable $XZ$, and the logical Z observable $ZZ$.
The Z basis rep code has the stabilizer $ZZ$, the logical X observable $XY$, and the logical Z observable $ZI$.
Beware that the choice of observables is slightly non-standard, and that their exact definition matters as it determines how errors propagate through the distillation process.
See \fig{distill_blocks} for circuits implementing these details correctly.

A rep code distillation is performed by Alice and Bob each measuring the stabilizer of the rep code, and comparing their results by using classical communication.
If the results differ from the result predicted by assuming they have shared copies of $|00\rangle + |11\rangle$, an error has been detected.
Equivalently, an error is detected if the stabilizer of the rep code anticommutes with the combination of the noisy Paulis applied to the input EPR pairs.
For example, an X basis rep code will raise a detection event if given a correct EPR pair and a phase flipped EPR pair.
Conversely, it won't detect a problem if given a bit flipped EPR pair or if given two phase flipped EPR pairs.

\begin{figure}
    \centering
    \resizebox{0.6\linewidth}{!}{
    \includegraphics{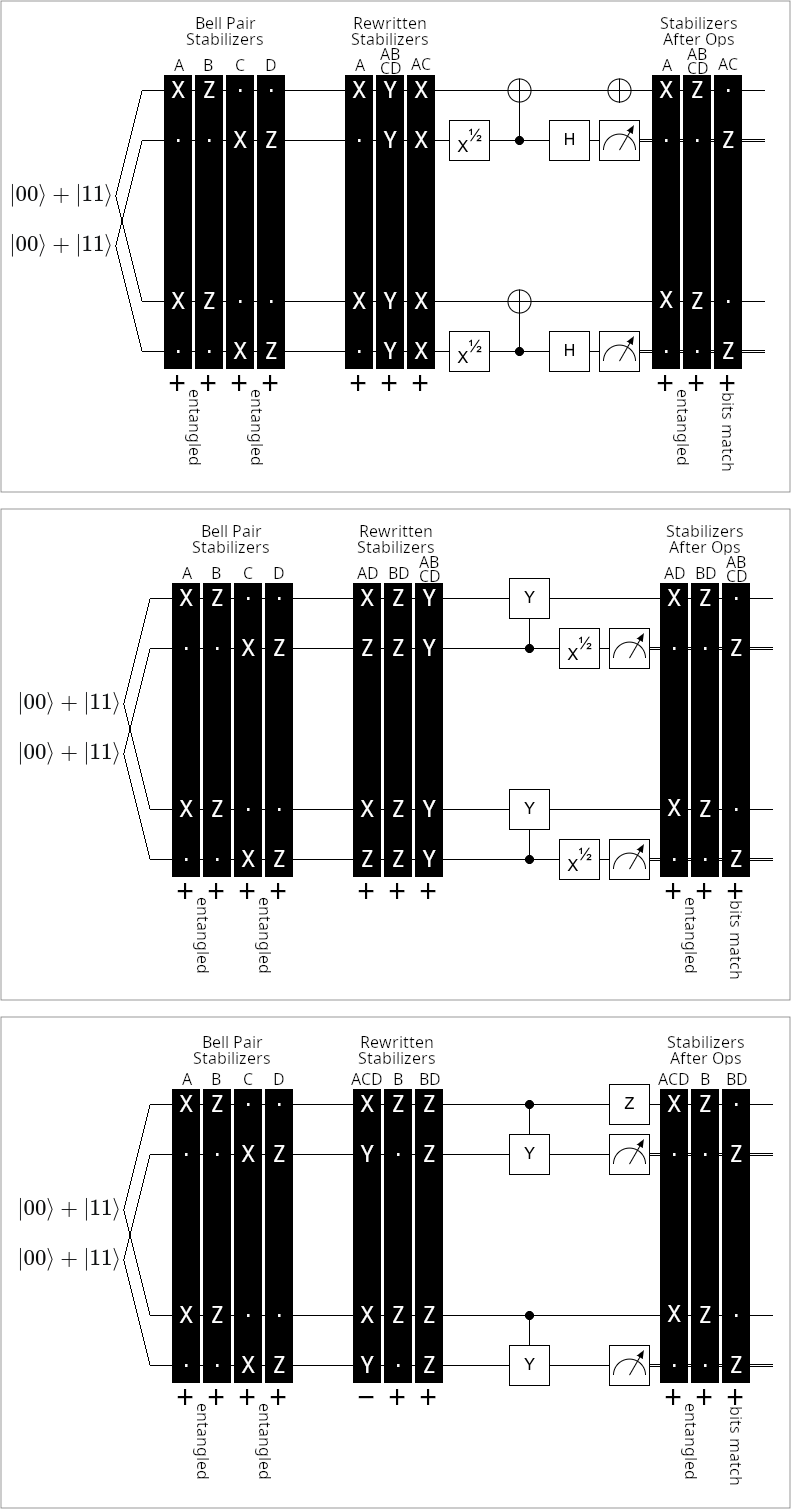}
    }
    \caption{
        How each rep code detects entanglement errors.
        The correct state $B_I$ has stabilizer generators $+XX$ and $+ZZ$.
        The error state $B_X$ negates the sign of $+ZZ$ to $-ZZ$.
        The error state $B_Z$ negates the sign of $+XX$ to $-XX$.
        The error state $B_Y$ negates the sign of both stabilizer generators.
        The pair of input EPR pairs has stabilizer generators $\pm XIXI$, $\pm ZIZI$, $\pm IXIX$, and $\pm IZIZ$.
        The signs are determined by the errors in the inputs.
        If the inputs are correct, then all signs are positive.
        For each rep code, there is a combination of the input stabilizers that is transformed by the operations into a $\pm XX$ stabilizer on the output qubits (and also a $\pm ZZ$ stabilizer).
        This is why the output qubits are still entangled.
        When the input states are correct, the output states have positive signs.
        When the input states have sign errors, these sign errors can propagate into the output states.
        Purification works by catching bad signs attempting to propagate into the output.
        Note how it's possible to combine the input stabilizers to form the stabilizer $+PPPP$, for each Pauli basis $P$.
        The basis $P$ rep code measures this stabilizer by having Alice locally measure $+PPII$ and Bob locally measure $+IIPP$.
        These local results can be combined using classical communication, to get the full measurement of $+PPPP$, allowing Alice and Bob to verify the state was in the +1 eigenstate of $+PPPP$ as expected.
        If one of the inputs was in a state other than $B_I$ or $B_P$, this catches the mistake.
        This filters out some errors that would otherwise propagate into the output, improving its entanglement fidelity.
    }
    \label{fig:distill_stabilizers}
\end{figure}

Stepping back, a rep code distillation takes two noisy EPR pairs as input.
It has some chance of detecting an error and discarding.
Otherwise it succeeds and produces one output.
See \fig{distill_stabilizers} for diagrams showing how each rep code catches errors.

To analyze rep code distillation, it's necessary to be able to compute the chance of discarding and the noise model describing the state output upon success.
I'll use $v \stackrel{B}{\star} w$ to represent the probability of a $B$-basis rep code distillation detecting an error.
I used sympy to enumerate the cases detected by each rep code and symbolically accumulate the probability of a failure.
The probabilities of a distillation failure for each basis are:

\begin{equation}
\begin{bmatrix}w_1\\x_1\\y_1\\z_1\end{bmatrix}
\stackrel{X}{\star}
\begin{bmatrix}w_2\\x_2\\y_2\\z_2\end{bmatrix}
= \frac{(w_1 + x_1)(y_2 + z_2) + (y_1 + z_1)(w_2 + x_2)}{(w_1+x_1+y_1+z_1)(w_2+x_2+y_2+z_2)}
\end{equation}

\begin{equation}
\begin{bmatrix}w_1\\x_1\\y_1\\z_1\end{bmatrix}
\stackrel{Y}{\star}
\begin{bmatrix}w_2\\x_2\\y_2\\z_2\end{bmatrix}
= \frac{(w_1 + y_1)(x_2 + z_2) + (x_1 + z_1)(w_2 + y_2)}{(w_1+x_1+y_1+z_1)(w_2+x_2+y_2+z_2)}
\end{equation}

\begin{equation}
\begin{bmatrix}w_1\\x_1\\y_1\\z_1\end{bmatrix}
\stackrel{Z}{\star}
\begin{bmatrix}w_2\\x_2\\y_2\\z_2\end{bmatrix}
= \frac{(w_1 + z_1)(x_2 + y_2) + (x_1 + y_1)(w_2 + z_2)}{(w_1+x_1+y_1+z_1)(w_2+x_2+y_2+z_2)}
\end{equation}

I'll use $u \stackrel{B}{\oplus} v$ to represent the state output by distilling a state $u$ against a state $v$ with a basis-$B$ rep code, given that no error was detected.
I used sympy to enumerate the cases not detected by each rep code, and symbolically accumulate the weight of each output case.
When distillation succeeds, the output states for each basis are:

\begin{equation}
\begin{bmatrix}w_1\\x_1\\y_1\\z_1\end{bmatrix}
\stackrel{X}{\oplus}
\begin{bmatrix}w_2\\x_2\\y_2\\z_2\end{bmatrix}
=\begin{bmatrix}
w_1w_2+x_1x_2\\
w_1x_2+x_1w_2\\
y_1y_2+z_1z_2\\
y_1z_2+z_1y_2\\
\end{bmatrix}
\end{equation}

\begin{equation}
\begin{bmatrix}w_1\\x_1\\y_1\\z_1\end{bmatrix}
\stackrel{Y}{\oplus}
\begin{bmatrix}w_2\\x_2\\y_2\\z_2\end{bmatrix}
=\begin{bmatrix}
w_1w_2+y_1y_2\\
x_1z_2+z_1x_2\\
w_1y_2+y_1w_2\\
x_1x_2+z_1z_2\\
\end{bmatrix}
\end{equation}

\begin{equation}
\begin{bmatrix}w_1\\x_1\\y_1\\z_1\end{bmatrix}
\stackrel{Z}{\oplus}
\begin{bmatrix}w_2\\x_2\\y_2\\z_2\end{bmatrix}
=\begin{bmatrix}
w_1w_2+z_1z_2\\
x_1y_2+y_1x_2\\
x_1x_2+y_1y_2\\
w_1z_2+z_1w_2\\
\end{bmatrix}
\end{equation}

Note that $u \stackrel{B}{\oplus} v$ is left-associative.
Also note that all these operators behave as you would expect with respect to decaying: if $a_1 \xrightarrow{\text{decay}} a_2$, and $b_1 \xrightarrow{\text{decay}} b_2$, and all states have less than 50\% infidelity, then $\left(a_1 \stackrel{B}{\oplus} b_1\right) \xrightarrow{\text{decay}} \left(a_2 \stackrel{B}{\oplus} b_2\right)$ and $\left(a_1 \stackrel{B}{\star} b_1\right) \leq \left(a_2 \stackrel{B}{\star} b_2\right)$.
This allows bounds on actual distilled states to be proved via bounds on decayed distilled states.

\subsection{Staging}

\begin{figure}
    \centering
    \resizebox{\linewidth}{!}{
    \includegraphics{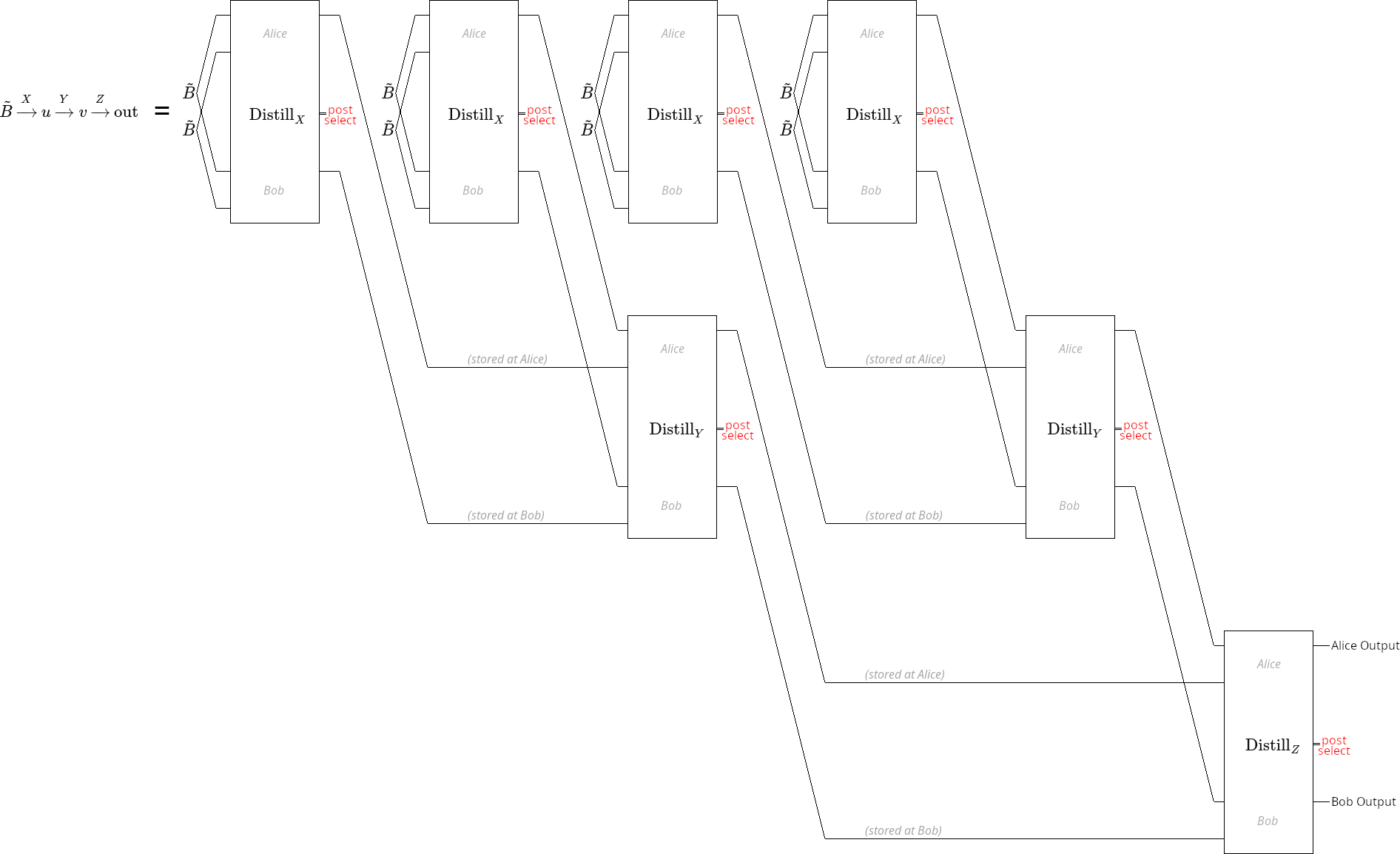}
    }
    \caption{
        A breakdown showing how three stages of rep code distillation are expanded into a circuit implementing the full distillation.
        Times moves from left to right.
        Stages and storage are laid out from top to bottom.
        Each stage's outputs are sequentially fed into the next stage as inputs.
        (Note that this diagram assumes no retries occur. If a retry did occur, the failing stage would discard its output and wait for more inputs to arrive in order to run again and produce a new potential output.)
        (Note that this diagram doesn't show what boosting looks like.
        Boosting would correspond to using the output of a stage as one of its inputs, some fixed number of times, before allowing the output to pass to the next stage.)
    }
    \label{fig:staging}
\end{figure}

Purifying with one enormous error detecting code doesn't work very well.
The issue is that, with large codes, the chances of seeing no detection events is extremely small.
This can result in over-discarding, where the purification process becomes inefficient because nothing is ever good enough.
It's more efficient to purify with a series of smaller stages, as shown in \fig{staging}, where each stage applies a small code and only successful outputs are collected to be used as inputs for the next stage.

For example, consider a process where input EPR pairs $u$ are distilled by an X basis rep code, and then survivor states $v$ are distilled by a Z basis rep code to produce final states $w$.
I describe this distillation chain using the notation $u \xrightarrow{X} v \xrightarrow{Z} w$.
The $[[4,1,2]]$ Shor code~\cite{shor1995qec} is also defined by concatenating a distance two X rep code with a distance two Z rep code, so you may expect a single stage distillation with the [[4,1,2]] code to behave identically.
The distillation chain $u \xrightarrow{[[4,1,2]]} w$ does produce the same output as $u \xrightarrow{X} v \xrightarrow{Z} w$, but it succeeds at a slightly slower rate.
The issue is that when the $[[4,1,2]]$ distillation detects a Z error, it costs 4 input pairs because the $[[4,1,2]]$ code takes all 4 data qubits in one group.
The two stage distillation detects Z errors during the first stage, so a Z error will only costs 2 input pairs.

A simple staging construction that purifies entanglement reasonably well is to alternate between stages that distill using the X basis rep code and stages that distill using the Z basis rep code:

$$\text{in} \xrightarrow{X} u_1 \xrightarrow{Z} u_2 \xrightarrow{X} u_3 \xrightarrow{Z} u_4 \xrightarrow{X} \dots \xrightarrow{X} u_{n} \xrightarrow{Z} \text{out}$$

The error suppression of an X stage followed by a Z stage is quadratic, because any single error will be caught by one stage or the other.
Furthermore, each stage only needs one qubit of storage because, as soon as two qubits are ready within one stage, they are distilled into one qubit in the next stage.
$O(1)$ additional stages cost $O(1)$ additional space to square the infidelity.
Therefore, given a target $\epsilon$, alternating between X rep code and Z rep code stages can reach that infidelity using $O(\log \log \frac{1}{\epsilon})$ storage.

\subsection{Boosting}

In the previous subsection, each stage distilled copies of the output from the previous stage.
Distillation was always combining two of the best-so-far state.
``Boosting'' is an alternative technique, where a good state is improved by repeatedly merging mediocre states into it.
(Note: in the literature, ``boosting'' is more commonly called ``pumping''~\cite{dr2003purification}).
In particular, staging as described in the previous subsection results in states being combined like this:

\begin{equation}
u_{k+3} = \left(\left(u_k \stackrel{Z}\oplus u_k\right) \stackrel{X}\oplus \left(u_k \stackrel{Z}\oplus u_k\right)\right) 
\stackrel{Z}\oplus
\left(\left(u_k \stackrel{Z}\oplus u_k\right) \stackrel{X}\oplus \left(u_k \stackrel{Z}\oplus u_k\right)\right)
\end{equation}

That kind of merging requires an amount of storage that increases with the amount of nesting.
You can instead combine states like this:

\begin{equation}
b_{k+1} = \left(\left(\left(\left(\left(\left(b_k \stackrel{Z}\oplus b_k\right) \stackrel{X}\oplus b_k\right) \stackrel{Z}\oplus b_k\right) \stackrel{X}\oplus b_k\right) \stackrel{Z}\oplus b_k\right) \stackrel{X}\oplus b_k\right) \stackrel{Z}\oplus b_k
\end{equation}

Here the result can be built up in a streaming fashion, with identical ``booster states'' arriving one by one to be folded into a single gradually improving ``boosted state''.

The upside of boosting is that it squeezes more benefit out of a state that you are able to repeatedly make.
The downside of boosting is that it can only be repeated a finite number of times.
The boosted state doesn't get arbitrarily good under the limit of infinite boosting.
It approaches a floor set by the booster state's error rates.
A second downside is that, because each booster state has a fixed chance of failing, the chance of detecting an error and having to restart the boosting process limits to 100\% as more boosts are performed.
The infidelity floor and the growing restart chance force you to eventually stop boosting and start a new stage.

\subsection{Bias Boosting Stage}

Suppose you have access to a booster state $A$ where the $X$, $Y$, and $Z$ errors of $A$ are all equal.
You are boosting a state where the $X$ and $Z$ terms are equal.
The following relationships hold:

\begin{equation}
\begin{bmatrix}1\\x\\y\\x\end{bmatrix}
\stackrel{Y}{\oplus}
\begin{bmatrix}1\\a\\a\\a\end{bmatrix}
=\begin{bmatrix}
1 + ay\\
2ax\\
y + a\\
2ax\\
\end{bmatrix}
\xrightarrow{\text{decay}}\begin{bmatrix}
1\\
2a x\\
y + a\\
2a x\\
\end{bmatrix}
\end{equation}

\begin{equation}
\begin{bmatrix}1\\x\\y\\x\end{bmatrix}
\stackrel{Y}{\star}
\begin{bmatrix}1\\a\\a\\a\end{bmatrix}
=\frac{2a(1 + y) + 2x(1 + a)}{(1+x+y+z)(1+a+a+a)}
\leq
2a + 2x
\end{equation}

In other words: boosting with $A$ using a $Y$ basis rep code additively increases the chance of a $Y$ error and the chance of discarding, but multiplicatively suppresses $X$ and $Z$ errors.

If this boosting process is repeated $\lfloor \sqrt[3]{1/a} \rfloor-1$ times, the output is:

\begin{equation}
\begin{bmatrix}1\\x\\y\\x\end{bmatrix}
\overbrace{
\stackrel{Y}{\oplus}
\begin{bmatrix}1\\a\\a\\a\end{bmatrix}
\stackrel{Y}{\oplus}
\begin{bmatrix}1\\a\\a\\a\end{bmatrix}
\stackrel{Y}{\oplus}
\dots
\stackrel{Y}{\oplus}
\begin{bmatrix}1\\a\\a\\a\end{bmatrix}
}^{\lfloor \sqrt[3]{1/a} \rfloor - 1\;\text{times}}
\xrightarrow{\text{decay}}\begin{bmatrix}
1\\
(2a)^{\lfloor \sqrt[3]{1/a} \rfloor - 1} x\\
y + a\lfloor \sqrt[3]{1/a} \rfloor - a\\
(2a)^{\lfloor \sqrt[3]{1/a} \rfloor - 1} x\\
\end{bmatrix}
\end{equation}

If the state being boosted is $A$ itself, then the result after all these boosts has exponentially smaller $X$ and $Z$ error rates (as long as the infidelity is small enough to begin with, e.g. less than 0.1\%):

\begin{equation}
\begin{bmatrix}1\\a\\a\\a\end{bmatrix}
\overbrace{
\stackrel{Y}{\oplus}
\begin{bmatrix}1\\a\\a\\a\end{bmatrix}
\stackrel{Y}{\oplus}
\begin{bmatrix}1\\a\\a\\a\end{bmatrix}
\stackrel{Y}{\oplus}
\dots
\stackrel{Y}{\oplus}
\begin{bmatrix}1\\a\\a\\a\end{bmatrix}
}^{\lfloor \sqrt[3]{1/a} \rfloor - 1\;\text{times}}
\xrightarrow{\text{decay}}\begin{bmatrix}
1\\
\frac{1}{2}(2a)^{\lfloor \sqrt[3]{1/a} \rfloor}\\
a\lfloor \sqrt[3]{1/a} \rfloor\\
\frac{1}{2}(2a)^{\lfloor \sqrt[3]{1/a} \rfloor}\\
\end{bmatrix}
\xrightarrow{\text{decay}}\begin{bmatrix}
1\\
\exp(-1/\sqrt[3]{a})\\
a^{2/3}\\
\exp(-1/\sqrt[3]{a})\\
\end{bmatrix}
\end{equation}

In this sequence of boosts, the first one is the most likely to discard.
The chance of any of the boosts discarding can be upper bounded by multiplying this probability by the number of boosts:

\begin{equation}
P(\text{bias boosting fails}) \leq (2a+2a) \cdot (\lfloor \sqrt[3]{1/a} \rfloor-1) \leq 4 a^{2/3}
\end{equation}

\subsection{Bias Busting Stage}

The bias boosting stage creates an enormous disparity between the $Y$ error rate and the other error rates.
The bias busting stage fixes this; bringing the $Y$ error rate down to match the others without increasing them by much.

This stage boosts using the biased state $B$ output from the previous stage.
This state has equal $X$ and $Z$ terms that are much smaller than its $Y$ term.
The boosts alternate between distilling with the X basis rep code and the Z basis rep code:

\begin{equation}
\begin{bmatrix}1\\x\\y\\x\end{bmatrix}
\stackrel{X}{\oplus}
\begin{bmatrix}1\\\alpha\\\beta\\\alpha\end{bmatrix}
\stackrel{Z}{\oplus}
\begin{bmatrix}1\\\alpha\\\beta\\\alpha\end{bmatrix}
=\begin{bmatrix}
1 + \alpha(x + x\beta + y\alpha)\\
x\beta + \alpha(\beta + y\beta + x\alpha)\\
y\beta^2 + \alpha(x + x\beta + \alpha)\\
x\beta + \alpha(1 + y + x\alpha)\\
\end{bmatrix}
\xrightarrow{\text{decay}}
\begin{bmatrix}
1\\
x\beta + 2\alpha\\
y\beta^2 + 2\alpha\\
x\beta + 2\alpha\\
\end{bmatrix}
\end{equation}

Assuming the initial infidelity is small enough (e.g. less than 0.1\%), the chance of this pair of boosts failing is at most:

\begin{equation}
\left(\begin{bmatrix}1\\x\\y\\x\end{bmatrix}
\stackrel{X}{\star}
\begin{bmatrix}1\\\alpha\\\beta\\\alpha\end{bmatrix}
\right)
+
\left(
\left(
\begin{bmatrix}1\\x\\y\\x\end{bmatrix}
\stackrel{X}{\oplus}
\begin{bmatrix}1\\\alpha\\\beta\\\alpha\end{bmatrix}
\right)
\stackrel{Z}{\star}
\begin{bmatrix}1\\\alpha\\\beta\\\alpha\end{bmatrix}
\right)
\leq
8x + 4y + 4\alpha + 4\beta 
\leq
10\beta
\end{equation}

The simplification to $10\beta$ is done by knowing that, immediately after a bias boosting stage, $\beta$ will be the largest term and the $\alpha$ and $x$ terms will be orders of magnitude smaller.

Repeating this pair of boosts $\lceil \frac{1}{2}\log_\beta(\alpha) \rceil$ times, starting from $B$, reduces all of the error terms below $4\alpha$:

\begin{equation}
\begin{bmatrix}1\\\alpha\\\beta\\\alpha\end{bmatrix}
\overbrace{
\stackrel{X}{\oplus}
\begin{bmatrix}1\\\alpha\\\beta\\\alpha\end{bmatrix}
\stackrel{Z}{\oplus}
\begin{bmatrix}1\\\alpha\\\beta\\\alpha\end{bmatrix}
\stackrel{X}{\oplus}
\begin{bmatrix}1\\\alpha\\\beta\\\alpha\end{bmatrix}
\stackrel{Z}{\oplus}
\begin{bmatrix}1\\\alpha\\\beta\\\alpha\end{bmatrix}
\stackrel{X}{\oplus}
\begin{bmatrix}1\\\alpha\\\beta\\\alpha\end{bmatrix}
\stackrel{Z}{\oplus}
\begin{bmatrix}1\\\alpha\\\beta\\\alpha\end{bmatrix}
}^{\lceil \frac{1}{2}\log_\beta(\alpha) \rceil\;\text{times}}
\xrightarrow{\text{decay}}
\begin{bmatrix}
1\\
4\alpha\\
4\alpha\\
4\alpha\\
\end{bmatrix}
\end{equation}

Since bias busting is always applied immediately after bias boosting, it will be the case that $\beta = a^{2/3}$ and $\alpha = \exp(-1/\sqrt[3]{a})$ for some $a$.
So the number of repetitions is at most $\frac{1}{2}\log_\beta(\alpha) \leq 1/\sqrt[3]{a}$.
The chance of discarding across all the bias busting boosts is therefore at most

\begin{equation}
P(\text{bias busting fails}) \leq 10\beta / \sqrt[3]{a} = 10a^{2/3} / \sqrt[3]{a} \leq 10 \sqrt[3]{a} = 10 \sqrt{\beta}
\end{equation}

Ensuring that $10\sqrt[3]{a}$ is less than 50\% requires starting from an $a$ below $0.01\%$.
You can start boosting from higher error rates $a$, without causing a disastrous discard rate, if you do fewer boosts in the bias boosting and bias busting stages.

\subsection{Full Construction}

An example of a full construction is shown in \fig{stages}.
The first step is to bootstrap from the initial infidelity to an infidelity low enough for the bias boosting and bias busting stages to start working.
Bootstrapping is done with unboosted rep code stages.
The bases of the rep codes are chosen by brute forcing a sequence that gets all error rates below 0.1\% in the fewest steps.

Once the error rate is bootstrapped, the construction begins alternating between bias boosting stages and bias busting stages.
The number of repetitions is customized during the first alternation, instead of using the loose bounds specified in previous sections, to get a faster takeoff.
Later alternations turn an infidelity of at most $a$ into an infidelity of at most $\exp(-1/\sqrt[3]{a})$.
This is not quite an exponential decrease, due to the cube root, but doing two alternations fixes this and guarantees a reduction to an infidelity below $\exp(-1/a)$.

Because each stage uses 1 qubit of storage, and a constant number of stages exponentiates the error rate, each additional exponentiation has $O(1)$ space cost.
It takes $O(\log^{\ast} \frac{1}{\epsilon})$ exponentiations to reach an infidelity of $\epsilon$, and therefore the storage needed to reach $\epsilon$ is $O(\log^{\ast} \frac{1}{\epsilon})$.

The last detail to discuss is the time complexity.
Because the discard rate at each stage is proportional to the infidelity at that stage, and the infidelity is decreasing so much from stage to stage, the sum of the discard rates across all stages converges to a constant as the number of stages limits to infinity.
This avoids one way that time costs can become non-linear versus number of stages.
That said, the input-to-output conversion efficiency is not quite asymptotically optimal.
Each stage is individually achieving a complexity of $\Theta(\log_{\epsilon_{\text{in}}} \frac{1}{\epsilon_\text{out}})$, with a constant factor of input-to-output waste.
As the number of stages increases, these waste factors compound.
As a result, the time complexity is $\Tilde{\Theta}(\log \frac{1}{\epsilon})$ instead of the optimal $\Theta(\log \frac{1}{\epsilon})$.

\begin{figure}
    \centering
        $$\begin{aligned}
        \xrightarrow{\text{in}}
        \begin{bmatrix}1\\1.7\cdot 10^{-1}\\1.7\cdot 10^{-1}\\1.7\cdot 10^{-1}\end{bmatrix}
        & \xrightarrow[\text{discards}\ 35\%]{\text{X}}
        \begin{bmatrix}1\\3.2\cdot 10^{-1}\\5.4\cdot 10^{-2}\\5.4\cdot 10^{-2}\end{bmatrix}
        \xrightarrow[\text{discards}\ 39\%]{\text{Y}}
        \begin{bmatrix}1\\3.5\cdot 10^{-2}\\1.1\cdot 10^{-1}\\1.1\cdot 10^{-1}\end{bmatrix}
        \xrightarrow[\text{discards}\ 29\%]{\text{X}}
        \begin{bmatrix}1\\7.0\cdot 10^{-2}\\2.3\cdot 10^{-2}\\2.3\cdot 10^{-2}\end{bmatrix}
        \\& \xrightarrow[\text{discards}\ 16\%]{\text{Y}}
        \begin{bmatrix}1\\3.2\cdot 10^{-3}\\4.6\cdot 10^{-2}\\5.4\cdot 10^{-3}\end{bmatrix}
        \xrightarrow[\text{discards}\ 9\%]{\text{Z}}
        \begin{bmatrix}1\\3.0\cdot 10^{-4}\\2.2\cdot 10^{-3}\\1.1\cdot 10^{-2}\end{bmatrix}
        \xrightarrow[\text{discards}\ 3\%]{\text{X}}
        \begin{bmatrix}1\\6.0\cdot 10^{-4}\\1.2\cdot 10^{-4}\\4.7\cdot 10^{-5}\end{bmatrix}
        \\& \xrightarrow[\text{discards}\ 2\%]{\text{bias (24 Y boosts)}}
        \begin{bmatrix}1\\10.0\cdot 10^{-81}\\3.0\cdot 10^{-3}\\9.6\cdot 10^{-81}\end{bmatrix}
        \xrightarrow[\text{discards}\ 10\%]{\text{bust (15 XZ boosts)}}
        \begin{bmatrix}1\\3.0\cdot 10^{-83}\\9.9\cdot 10^{-79}\\9.6\cdot 10^{-81}\end{bmatrix}
        \\&\xrightarrow[\text{discards}\ <1\%]{\text{bias (}10^{27}\text{ Y boosts)}}
        \begin{bmatrix}1\\10^{-10^{28}}\\10^{-55}\\10^{-10^{28}}\end{bmatrix}
        \xrightarrow[\text{discards}\ <1\%]{\text{bust (}10^{27}\text{ XZ boosts)}}
        \begin{bmatrix}1\\10^{-10^{28}}\\10^{-10^{28}}\\10^{-10^{28}}\end{bmatrix}
        \end{aligned}$$
    \caption{
        An example 10-stage sequence that purifies entanglement with an infidelity of 1/3 into entanglement with an infidelity of $10^{-10000000000000000000000000000}$.
        Error rates and discard rates were computed using a python script, except for the last two stages.
        They underflowed standard floating point numbers and so had to be estimated based on bounds described elsewhere in the paper.
    }
    \label{fig:stages}
\end{figure}

\section{Conclusion}
\label{sec:conclusion}

In this paper, I showed that surprisingly little storage is needed to purify entanglement, assuming storage and local operations and classical communication are noiseless.
An infidelity of $\epsilon$ can be reached using $O(\log^\ast \frac{1}{\epsilon})$ storage and $\Tilde{O}(\log \frac{1}{\epsilon})$ time.
Although I focused on an asymptotic limit more interesting to theory than practice, I'm hopeful that the presented ideas will inspire practical constructions.

\printbibliography

\end{document}